\documentstyle[aps,prb]{revtex}

\begin{document}

\title{Absence of two energy scales in the two-impurity Kondo Model} 
\author{Kurt Fischer\cite{Fischer-address-Tokyo}}
\address{Max-Planck-Institut f\"ur Physik komplexer Systeme, N\"othnitzer Strasse 38, 01187 Dresden, Germany} 
\date{December 24th}
\maketitle

\begin{abstract}
It is believed that the successive antiferromagnetic scattering of the
conduction electrons on two magnetic impurities in a metal, induces a magnetic
interaction between the impurities which sets an energy scale for the system,
in addition to the Kondo temperature.
However, it is shown here that this contradicts a scaling law for the 
two-impurity Anderson model in the magnetic limit which becomes exact for a
large bandwidth. 

\end{abstract}


In heavy fermion materials localized, strongly correlated $f$ electrons
interact with delocalized, weakly correlated conduction electrons.  
Experiments show that above a temperature $T^*$ the magnetic moments of
the $f$ electrons behave asymptotically as free. 

Well below $T^*$ they can either be phenomenologically described as a Fermi
liquid, or magnetic correlations between the local $f$ moments become
dominant.~\cite{review-Grewe/Steglich} 

When searching for a microscopic mechanism, one is guided by the behavior of
dilute magnetic alloys which are well described by the single-impurity
Anderson Hamiltonian.~\cite{Buch-Hewson}
At low temperatures the magnetic moment of the impurity is screened by
particle-hole excitations of the conduction band.
The system behaves universally, that is, all observables scale with one energy
$k_B T_K$ where $T_K$ is called the Kondo temperature.
On the other hand, the heavy fermion materials have, in orders of magnitude, 
one magnetic ``impurity'' atom per unit cell.
The hypothesis for such systems is that it can be described by the
Anderson-lattice Hamiltonian.  

It is far from obvious why the single-impurity Kondo effect should survive in
the lattice model.
It is believed that magnetic interactions between the local $f$ moments
compete with the Kondo effect.  
This magnetic interaction between the local $f$ moments is supposed to be
induced by successive scattering of the conduction electrons on the two
impurities, and called Ruderman-Kittel or RKKY
interaction.~\cite{Ruderman/Kittel,Doniach-two-imp,Jayaprakash/Krishna-murthy/Wilkins-two-imp}
Doniach~\cite{Doniach-two-imp} estimated how much energy the
system gains if it orders antiferromagnetically.
By second-order perturbation theory in the exchange coupling $J$ between the
conduction and the $f$ electrons this RKKY energy was found to be of the order 
\[
I \propto \rho^2 J^2 D ,
\]
where $\rho$ is the density of states at the Fermi energy of the conduction
band of width $D$.  
Doniach compared it with the exponentially small energy gain 
\[
k_B T_K \propto D \exp [ - 1/(\rho J) ] ,
\]
if the magnetic atoms are considered to be independent magnetic impurities.
Depending on the size of $J$, it could then be estimated whether the material
is in a magnetic or nonmagnetic phase, at low temperatures. 
This simple picture deals essentially with a two-impurity Hamiltonian, the
standard version of which is the two-impurity Anderson or Kondo model.
It has had great impact on the experimentalist's point of view. 
In fact, standard reviews quote the Doniach picture to explain, at least
qualitatively, the respective experimental 
observations.~\cite{review-Grewe/Steglich,review-Fulde/Keller/Zwicknagl}  

In this article it will be shown that this simple picture cannot be valid. 
To this end let us discuss firstly the various approaches to the
two-impurity problem.   

For the two-impurity Kondo model, summing of leading-order divergent
diagrams~\cite{Varma-two-imp} in the spirit of
Abrikosov~\cite{Abrikosov-Kondo} showed that at decreasing temperature, the
RKKY interaction as well as the Kondo interaction are renormalized, and
eventually diverge at low temperatures.   
Hence the picture drawn by Doniach would be invalid.

This can be contrasted with results from poor man's
scaling:~\cite{Jayaprakash/Krishna-murthy/Wilkins-two-imp,Krishna-murthy/Jayaprakash-two-imp}
Depending on the distance between the impurities, the 
RKKY interaction $I$ may either be ferro- or antiferromagnetic.
If $I$ is ferromagnetic and larger than $k_B T_K$, it is predicted that the
magnetic moments of the two impurities first form an effective spin one and
are then quenched in a two-stage process, as temperature decreases. 
On the other hand, a large antiferromagnetic RKKY interaction suppresses the
Kondo effect and the two impurity spins form a singlet for $k_B T < I$.
It is then argued that $I$ sets an energy scale for the system at which the
RKKY interaction could be observed.

This analysis was supported by a more recent calculation with the help of a
variant of the numerical renormalization group by Silva {\it et
al.}~\cite{Silva/Lima/Oliveira/Mello/Oliveira/Wilkins-two-imp,Oliveira/Oliveira-two-imp}
(in which the $N$th hopping matrix element of the half-infinite chain to
which the two impurities couple is vanishing as 
$\Lambda^{-N}$ with $\Lambda  =  10$, and subsequently the results of
different band-widths 
$D / \Lambda^\epsilon, \epsilon  =  0 \dots 1$
are averaged over). 
In fact, those results imply that the Doniach picture is {\it
valid}. 

However, those results contradict others~\cite{Sakai/Shimizu-two-imp-1}
using the conventional renormalization group scheme ($\Lambda = 3$) where
only one low-energy scale was seen.  
What is more, the approaches of
Refs.~\onlinecite{Jayaprakash/Krishna-murthy/Wilkins-two-imp,Krishna-murthy/Jayaprakash-two-imp}, and~\onlinecite{Silva/Lima/Oliveira/Mello/Oliveira/Wilkins-two-imp} are inconsistent. 
To clarify the point, only the case of an antiferromagnetic RKKY interaction
is discussed.
Silva {\it et al.} interpreted their data with the help of poor man's scaling.
They saw that the magnetic moment of the impurities is quenched
as temperature decreases below $I/k_B$, at which temperature $I$ and $J$
themselves are not much renormalized. 
They claimed that therefore the two impurities would form a
singlet for $k_B T<I$, consequently the renormalization of $J$ would stop,
and the low-temperature behavior could be determined from perturbation
theory, the lowest nonvanishing order of which is $\propto J^2/I$.
A simple estimation shows however that this perturbative expansion is not
allowed:
Because $\rho\propto 1/D$, the RKKY interaction $I$ is always a factor $J/D$
{\it smaller} then the Kondo interaction $J$.
This can be read off their numerical
results as well.~\cite{Silva/Lima/Oliveira/Mello/Oliveira/Wilkins-two-imp}

Incidentally, if the arguments given by Silva {\it et al.} were
correct they could be repeated for the antiferromagnetic single-impurity
Kondo model:
At a temperature of the order of the coupling constant $J$ the impurity spin
and an electron of the conduction band would form a singlet,
the renormalization would stop, and one would be left with a ``Kondo effect''
at an energy scale of the order of $J$, which is certainly not true.

To summarize, the Doniach picture and its subsequent refinements as
described above, imply that the renormalization of $J$ stops as
temperature becomes lower than the RKKY coupling. 
This RKKY interaction then sets an energy scale for the system's response, at
least for small coupling $J \ll D$.  
Therefore it would be helpful if the exact energy scales of a
metal with two magnetic impurities could be calculated in the universal limit 
of large $D$.
The method which is usually employed for that purpose is the diagrammatic
renormalization group.  
The major obstacle is the solution of the Dyson equation connecting the
dressed propagators and their self-energies which can be done only
perturbatively.  
However, the low-energy behavior is beyond perturbation theory.
An explicit solution of the Dyson equation can be
avoided~\cite{Fischer-scaling-imp} by means of the variational principle of 
Luttinger and Ward~\cite{Luttinger/Ward,KuramotoI,Baym} with the help of which
the energy scales will follow.

The two-impurity Anderson model serves here as the standard model
for two magnetic impurities in a metal. 
Mean-field approaches for this model have the drawback that the RKKY
interaction appears only as a higher-order
correction,~\cite{Jones/Kotliar/Millis} and therefore a magnetic interaction 
between the impurities had to be added by hand, giving a model with an 
adjustable strength of the magnetic exchange between the impurities.

Here, however, a direct exchange interaction between the impurities will 
{\it not} be taken into account, because the two impurities model two 
localized $f$ orbitals on different lattice sites, with virtually no overlap,
and it is the purpose of this article to investigate the role of their 
{\em induced} magnetic interaction, that is, the RKKY interaction.

The two impurities are assumed to sit at $(0,0,\pm R/2)$ symmetrically to the
origin of the coordinate system, along the $z$ axis.
The Hamiltonian $H$ is then invariant under the parity transformation. 
Hence the creation operator for a conduction electron of momentum $p$ and
spin $m$ and for an impurity electron can be expressed in terms of their even
($\sigma =  1 $) and odd ($\sigma =  -1 $)
combinations,~\cite{Coleman-two-imp,Schiller/Zevin-two-imp-1} 
\begin{eqnarray}\label{slave-particles-Def}
c_{\vec{p},m,\sigma}^+    
&=&     \frac{1}{\sqrt{2}} \left( 
        c_{\vec{p},m}^+ + \sigma c_{-\vec{p},m}^+ \right) , \nonumber \\
f_{\sigma m}^+ |  \text{vac}  \rangle 
&=&     \frac{1}{\sqrt{2}} \left( 
        f_{1m}^+ + \sigma f_{2m}^+ \right)|  \text{vac}  \rangle        , \\
        d_{\sigma mn}^+|  \text{vac}  \rangle 
&=&     - \sigma d_{\sigma nm}^+|  \text{vac}  \rangle 
=       \frac{\left( 
        \sigma f_{1m}^+ f_{2n}^+ - f_{1n}^+ f_{2m}^+ 
        \right)}
        {\sqrt{ 2(1+\delta_{mn}) }} |  \text{vac}  \rangle  . \nonumber
\end{eqnarray}
Here  $c_{\vec{p}m}^+$ creates a conduction electron with internal quantum
number $m = 1 \dots N$, momentum $\vec{p}$, and energy $\epsilon(\vec{p})$.
If $|  \text{vac}  \rangle$ is the vacuum, then 
$b^+ |  \text{vac}  \rangle$, $f_{im}^+ |  \text{vac}  \rangle$, and 
$d_{mn}^+|  \text{vac}  \rangle$ 
denote the unoccupied, singly occupied, and doubly occupied impurity
configurations, with $i = 1, 2$ labeling the impurities. 
The energy difference between the singly occupied and unoccupied as well as
between the doubly and singly occupied impurity configuration is $\epsilon_f$.
$f_{im}^+$ is a fermionic operator, and $b^+$ and $d_{mn}^+$ are bosonic.
Double occupancy at each impurity site is suppressed by requiring
$ b^+b + \sum_{\sigma m} f_{\sigma m}^+f_{\sigma m}^{\phantom{+}} 
+ \sum_{\sigma, m\geq n} d_{\sigma mn}^+ d_{\sigma mn}^{\phantom{+}} \equiv 1
$. 
The unoccupied configuration $ b^+ |  \text{vac}  \rangle$ has even parity.
The effective Hamiltonian then reads,~\cite{Schiller/Zevin-two-imp-1}
\begin{eqnarray}
H   &=& H_c + H_f + H_1                                 ,       \nonumber \\
H_c &=& \sum_{\epsilon\sigma} \epsilon 
        c_{\epsilon m\sigma}^+ c_{\epsilon m\sigma}^{\phantom{+}},\nonumber \\ 
H_f &=& \epsilon_f \sum_{m\sigma} f_{\sigma m}^+f_{\sigma m}^{\phantom{+}} 
        + 2 \epsilon_f \sum_{m\geq n , \sigma} 
        d_{\sigma mn}^+ d_{\sigma mn}^{\phantom{+}}     ,       \nonumber \\
H_1 &=& \frac{1}{\sqrt{N}} \sum_{\epsilon,m,\sigma}   V_\sigma (\epsilon)
        \left( f_{\sigma m}^+  b \ c_{\sigma \epsilon m}^{\phantom{+}}
        + H.c. \right)                                          \nonumber \\
&&      + \frac{1}{\sqrt{2N}} \sum_{\epsilon,m\neq n, \sigma,\tau}  
        \tau V_\tau (\epsilon)
        \left( d_{\sigma \tau, mn}^+  f_{\sigma n}^{\phantom{+}} 
        c_{\tau \epsilon m}^{\phantom{+}}  + H.c. \right)       \nonumber \\
&&      + \frac{1}{\sqrt{N}} \sum_{\epsilon,m,\sigma} 
        \sigma V_\sigma (\epsilon)
        \left( d_{-1, mm}^+  f_{-\sigma m}^{\phantom{+}} 
         c_{\sigma, \epsilon m}^{\phantom{+}} + H.c. \right)  .
\end{eqnarray}
The impurity hybridization with the conduction band is proportional to $V$. 
The Boltzmann constant is set to unity so that temperature is
measured in units of energy.
The dispersion can be assumed as isotropic.  
The density of states $\rho$ is assumed to be constant, and the 
such linearized dispersion is cut off at $\pm D$.
The effective hybridization matrix element can then be expressed in terms of
the conduction band density of states as~\cite{Schiller/Zevin-two-imp-1} 
\begin{equation}\label{V-scaling}
V_\sigma(\epsilon) = 
V \sqrt{\rho} \Theta(1 - \frac{\epsilon^2}{D^2})
\sqrt{1 + \sigma \frac{ \sin [  k_F R  ( 1+ \epsilon / 2D ) ] }
                        { k_F R( 1+ \epsilon / 2D )} ,
} 
\end{equation}
with $k_F$ the Fermi wave number and $\Theta$ the step function.
Thus the system has been reduced to an effective one-impurity
Hamiltonian, the additional parity quantum number keeping track of the
original two impurities. 
  
The one-particle impurity propagators $R_f$ for the unoccupied ($f=0$),
singly occupied ($f=\sigma m$), and doubly occupied ($f=\sigma mn$)
impurity configurations, determine the impurity part $Z_f$ of the partition
function via a line integral, the path of integration encircling all poles of
the integrand  
\begin{equation}\label{Z-representation}
Z_f := \frac{ \text{Tr}_f \text{Tr}_c e^{-\beta H} }{ \text{Tr}_c e^{-\beta
        H_c} }  
= \text{Tr}_f \oint \frac{dz}{2\pi i} e^{-\beta z} R_f(z)  , 
\end{equation}
where $Tr_f$ denotes the trace over the impurity configurations.
The propagators $R_f$ can be calculated by the standard diagrammatic
technique~\cite{Schiller/Zevin-two-imp-1} which is reviewed in
Ref.~\onlinecite{review-Bickers}. 
The vertices and naked propagators are shown in
Fig.~\ref{Vertices-Anderson-two-imp}.     

Within the variational principle,~\cite{KuramotoI} a functional $\Upsilon$ of
the propagators $R_f$ is defined in terms of skeleton diagrams. 
For the impurity part of the partition function,  $\Upsilon$ is given by  
\begin{eqnarray}\label{Upsilon}
\Upsilon &=& \beta \text{Tr}_f \oint \frac{dz}{2 \pi i }  e^{-\beta z}   
         \big\{ \sum_n
         (1-\frac{1}{n}) \Sigma_f^{(n)}(R_f(z)) R_f(z) 
         \nonumber \\
&& +     \ln [ z-H_f - \sum_n \Sigma_f^{(n)}(R_f(z)) ] \big\} . 
\end{eqnarray}
Here $\Sigma_f^{(n)}$ denotes the $n$th order self-energy of $R_f$, expressed 
in terms of skeleton diagrams.
At the saddle point with respect to variations of $R_f$, the functional
$\Upsilon$ equals $Z_f$ and the Dyson equation holds as a
self-consistency equation.~\cite{KuramotoI}
So $\Upsilon$ depends on parameters such as $\epsilon_f$ explicitly only via
$H_f$, and not implicitly via the propagators.~\cite{KuramotoI}  
In addition, $\Upsilon$ depends explicitly on $V$ only via the prefactor
$V^{2n}$ of the $2n$th order self-energy $\Sigma_f^{(2n)}$.     
To determine the $T$ dependence, the internal integration
variables $z$ and $\epsilon$, as in
Eqs.~(\ref{Upsilon}) and (\ref{NCA-equations-two-imp}) are replaced by $Tz$,
$T\epsilon$, and $D/T$, respectively.     
The variational principle remains unaffected.
$\Upsilon$ depends now explicitly on $T$ via the prefactor $T^n$ of a skeleton
diagram of $2n$th order, the term $Tz$ in the logarithm, and $D/T$ because of
Eq.~(\ref{V-scaling}). 
Similar to Ref.~\onlinecite{Fischer-scaling-imp} it follows for the
impurity part $F_f$ of the free energy
\begin{equation}\label{F-scaling-first-two-imp}
F_f =   \left( 
        T \frac{\partial}{\partial T} 
        + \rho V^2  \frac{\partial}{\partial \rho V^2}
        + \epsilon_f  \frac{\partial}{\partial \epsilon_f} 
        + D  \frac{\partial}{\partial D} 
         \right)  F_f  . 
\end{equation}
Approximations fulfilling the variational principle can be generated
by using a subclass, the so-called families of
skeleton diagrams.~\cite{Grewe1983}  
If, for instance, only the family of second-order skeleton diagrams is kept in
$\Upsilon$, the self-energies are ($f$ denotes the Fermi function),
\begin{eqnarray}\label{NCA-equations-two-imp}
\Sigma_{\sigma\tau,mn}^{(2)}(z) 
&=&     \frac{ 1 }{2N}  \sum_{\sigma\tau = const} 
        \int d\epsilon f(\epsilon) V^2_\tau (-\epsilon)  
        [ R_{\sigma n} (z+\epsilon)                             \nonumber \\
&&      + R_{\sigma m} (z+\epsilon) ] , \nonumber \\
\Sigma_{\sigma m}^{(2)}(z)                                              
&=&     \frac{1}{N}  \int  d\epsilon  f(\epsilon)
        \big[ \frac{1}{2} \sum_{\tau,n\neq m} V^2_\tau (\epsilon) 
        R_{\sigma\tau,mn} (z+\epsilon)                          \nonumber \\
\lefteqn{ + V^2_\sigma(-\epsilon) R_0(z+\epsilon)
        + V^2_{-\sigma}(\epsilon) R_{-1,mm}(z+\epsilon) 
        \big] ,         }                                       \nonumber \\
\Sigma_0^{(2)}(z) 
&=&     \frac{1}{N} \sum_{\sigma,m} \int  d\epsilon f(\epsilon)  
        V_\sigma^2(\epsilon)  R_{\sigma m}(z+\epsilon)   .   
\end{eqnarray}
Universal behavior of the system manifests itself when $D$ becomes larger
than all other energy scales of the system.  
$\Upsilon$ depends explicitly on $D$ only via $V_\sigma^2$ as in
Eq.~(\ref{V-scaling}),
\begin{equation}\label{F-D-derivative}
D\partial_D \Upsilon = 
-  \text{Tr}_f \oint \frac{\beta dz}{2 \pi i }  e^{-\beta z}   
D\partial_D \sum_n \frac{1}{n} \Sigma_f^{(n)}(z) R_f(z)  . 
\end{equation}
The skeleton diagrams for $\Sigma_f^{(2)} R_f$ are shown in
Fig.~\ref{NCA-diagrams-two-imp}.       

The effective density of states of Eq.~(\ref{V-scaling}) scales with $D$ as a
function of $\epsilon$ and is finite at the Fermi energy $\epsilon=0$.
Hence,~\cite{Fischer-scaling-imp} as in the case for a density of states
with a sharp cutoff at $\pm D$, differentiating a skeleton diagram for
$\Sigma_f^{(n)} R_f$ with respect to $D$ amounts to removing one curved
conduction electron line and replacing the internal propagator by its value at
the cutoff $\propto  1/D$.
Therefore a skeleton diagram of second order contributes $\propto  D/D$ to
the logarithmic derivative. 

A skeleton diagram of higher than second order contains vertex corrections;
hence there lie under each conduction electron line at least two
impurity propagators, because otherwise this diagram would have a self-energy 
insertion and not be a skeleton.
Its contribution to the logarithmic derivative is therefore $\propto  D/D^2$
and can be neglected for large $D$.~\cite{Fischer-scaling-imp}

Therefore, Eq.~(\ref{F-D-derivative}) only the skeleton diagrams for
$\Sigma_f^{(2)} R_f$ are necessary to obtain the exact energy scales of the
system in the universal limit of large $D$.
A spectral decomposition of the impurity propagators yields,
\begin{eqnarray*}
D \frac{\partial}{\partial D} F_f 
&=&     - \int \frac{e^{-\beta\omega} d\omega }{ N Z_f } 
        \bigg\{ \sum_{\sigma, m} V_\sigma^2(0) 
        \left[ \rho_{\sigma m}(\omega) + \rho_0(\omega) \right]         \\
&&      + \frac{1}{2}  \sum_{\sigma\tau, m\neq n}  
          V_\tau^2(0)  \left[ \rho_{\sigma\tau mn}(\omega) 
        + \rho_{\sigma n}(\omega) \right]                               \\
&&      + \sum_{\sigma, m} \left[ 
        V_\sigma^2(0)  \rho_{-1,mm}(\omega)  
        + V_{-\sigma}^2(0)  \rho_{\sigma m}(\omega) \right]
        \bigg\}  . 
\end{eqnarray*}
Together with the identities 
\[
Z_f  =  \int e^{-\beta\omega} d\omega 
        [ \rho_0(\omega) + \sum_{\sigma, m} \rho_{\sigma m}(\omega) 
        + \sum_{\sigma, m\geq n} \rho_{\sigma mn}(\omega) ]     ,
\]
\[
Z_f  \frac{\partial}{\partial \epsilon_f} F_f                   
=       \int e^{-\beta\omega} d\omega 
        [ \sum_{\sigma, m} \rho_{\sigma m}(\omega) 
+       2 \sum_{\sigma, m\geq n}  \rho_{\sigma mn}(\omega) ] ,
\]
the following scaling equation is obtained,
\begin{equation}\label{F-scaling-second-two-imp}
D \frac{\partial}{\partial D} F_f =  
        \rho V^2  \left( 1- 1/N \right)
        \frac{\partial}{\partial \epsilon_f} F_f - 2 \rho V^2  .
\end{equation}
Eqs.~(\ref{F-scaling-first-two-imp}) and~(\ref{F-scaling-second-two-imp}) 
imply the scaling law  
\begin{equation}\label{F-scaling-two-imp}
F_f - E_0 = 
T g(T,V,\epsilon_f,D, \rho) =  
T g \left( \frac{T}{\Gamma} , \frac{T}{T_K} \right) 
\end{equation} 
in the magnetic limit $ -\epsilon_f \gg \rho V^2$, with 
$\Gamma \! = \! \pi \rho V^2/N$.
In this limit, the Anderson model is equivalent to the two-impurity
Kondo model~\cite{Schrieffer-Wolff} at low temperatures $T \ll \rho V^2$,
with $T_K =  D \sqrt[N]{ \rho V^2 / D } \exp\left[  \epsilon_f/ \rho V^2
\right] $.
By the same means~\cite{Fischer-scaling-imp} it can be shown that in this 
limit all observables scale as in Eq.~(\ref{F-scaling-two-imp}).

Hence the two-impurity Kondo model has in the limit of large $D$ only 
{\it one} low energy scale, which is proportional to the single-impurity Kondo 
temperature.
Observe, however, that the proportionality factor cannot be fixed within this
approach and will of course depend on $k_F R$.

If the RKKY interaction would set additional energy scales as discussed  in
the introduction, then the scaling law would have the form,
\begin{equation}\label{F-scaling-Tk-RKKY}
F_f -E_0 = g \left( \frac{T}{T_K} , \frac{T}{I} , \frac{T}{T_{KA}} ,
\frac{T}{T_+} \right) 
\end{equation}
where $T_{KA}$ and $T_+$ are the temperatures of the two-stage Kondo effect for
the ferromagnetic
RKKY interaction.~\cite{Jayaprakash/Krishna-murthy/Wilkins-two-imp,Krishna-murthy/Jayaprakash-two-imp}
This, however, contradicts the scaling law~(\ref{F-scaling-two-imp}) above.

To conclude, if the high-energy cutoff of the metal is so large that terms of 
the order $T/D$ or $J/D$ can be neglected, there is for two magnetic 
impurities in a metal only one low-energy scale.
The induced indirect exchange interaction between the local $f$ moments which 
is mediated by successive scattering of conduction electrons, will be strongly
renormalized at low temperatures.
However, the scaling equations alone do not predict the nature of the ground
state for two magnetic impurities in a metal.
In view of Eq.~(\ref{F-scaling-two-imp}) it seems doubtful whether a
competition between the formation of magnetic order and a Fermi liquid of heavy
quasiparticles can be understood without adding explicitly a magnetic 
interaction between the magnetic $f$ moments.
  
While preparing this work the author has benefited from numerous discussions
with Tom Schork, Karen Hallberg, Karlo Penc, and Professor Peter Fulde.




\begin{figure}
\caption{\label{Vertices-Anderson-two-imp}
Vertices for the two-impurity Anderson model. 
A dashed, singly or doubly arrowed line represents the naked propagator
of the singly or doubly occupied impurity configuration.
A single $\sigma$ or $\tau$ labels the parity $\pm 1$, and $\sigma\tau$ their
product. 
$m$ or $n$ labels the magnetic quantum number, and $mn$ a pair of
magnetic quantum numbers for the double occupied configuration. 
A wavy respective solid line represents the naked propagator of the unoccupied
impurity configuration or the conduction electron. 
The product of parities of incoming particles equals that of the outgoing
ones.}
\end{figure}

\begin{figure}
\caption{\label{NCA-diagrams-two-imp}
Skeleton diagrams of second order for the two-impurity Anderson model}
\end{figure}

\end{document}